\documentclass[aps,prb,twocolumn,superscriptaddress,notitlepage]{revtex4-1}
\usepackage{graphicx}
\usepackage[caption=false]{subfig}
\usepackage{float}
\usepackage{natbib}
\usepackage{xcolor}
\usepackage{xr}
\usepackage{amsmath}
\usepackage{amssymb}
\usepackage{array}
\usepackage{wasysym}
\usepackage{color,soul}
\usepackage{braket}
\usepackage{verbatim}
\usepackage{hyperref}
\usepackage{gensymb}
\usepackage{url}
\usepackage{dirtytalk}
\usepackage{upgreek}

\usepackage{filecontents}
\begin{document}

\title{Uniaxial stress enhanced anisotropic magnetoresistance and superconductivity in the kagome superconductor LaRu$_{3}$Si$_{2}$}

\author{P. Král}
\thanks{These authors contributed equally.}
\affiliation{PSI Center for Neutron and Muon Sciences CNM, 5232 Villigen PSI, Switzerland}

\author{V. Sazgari}
\thanks{These authors contributed equally.}
\affiliation{PSI Center for Neutron and Muon Sciences CNM, 5232 Villigen PSI, Switzerland}

\author{Yongheng Ge}
\thanks{These authors contributed equally.}
\affiliation{Wuhan National High Magnetic Field Center and School of Physics, Huazhong University of Science and Technology, Wuhan 430074, China}

\author{O. Gerguri}
\affiliation{PSI Center for Neutron and Muon Sciences CNM, 5232 Villigen PSI, Switzerland}

\author{M. Spitaler}
\affiliation{PSI Center for Neutron and Muon Sciences CNM, 5232 Villigen PSI, Switzerland}

\author{J.N. Graham}
\affiliation{PSI Center for Neutron and Muon Sciences CNM, 5232 Villigen PSI, Switzerland}

\author{H.~Nakamura}
\affiliation{Institute for Solid State Physics (ISSP), University of Tokyo, Kashiwa, Chiba 277-8581, Japan}

\author{M. Bartkowiak}
\affiliation{PSI Center for Neutron and Muon Sciences CNM, 5232 Villigen PSI, Switzerland}

\author{S.~Nakatsuji}
\affiliation{Institute for Solid State Physics (ISSP), University of Tokyo, Kashiwa, Chiba 277-8581, Japan}

\author{H. Luetkens}
\affiliation{PSI Center for Neutron and Muon Sciences CNM, 5232 Villigen PSI, Switzerland}

\author{G. Simutis}
\affiliation{PSI Center for Neutron and Muon Sciences CNM, 5232 Villigen PSI, Switzerland}

\author{Gang~Xu}
\email{gangxu@hust.edu.cn} 
\affiliation{Wuhan National High Magnetic Field Center and School of Physics, Huazhong University of Science and Technology, Wuhan 430074, China}

\author{Z. Guguchia}
\email{zurab.guguchia@psi.ch}
\affiliation{PSI Center for Neutron and Muon Sciences CNM, 5232 Villigen PSI, Switzerland}
\date{\today}

\begin{abstract}
Elucidating the role of the kagome electronic structure in determining the various quantum ground states is of fundamental importance. In this work, we employ in-plane uniaxial stress as a tuning parameter to probe the electronic structure and its impact on the superconducting and normal-state properties of the kagome superconductor LaRu$_{3}$Si$_{2}$, combining magnetotransport measurements with first-principles calculations. We identify a pronounced anisotropy in both the upper critical field and the normal-state magnetoresistance, indicating strong electronic anisotropy despite the three-dimensional crystal structure. Furthermore, we find that the superconducting transition temperature $T_{\rm c}$ increases under in-plane stress applied within the kagome plane, although the enhancement is modest, reaching approximately 0.3 K at 0.6 GPa. Furthermore, the absolute magnetoresistance exhibits a pronounced increase from about 22${\%}$ at zero stress to 35${\%}$ at 0.6 GPa, indicating a substantial modification of the normal state above $T_{\rm c}$. Previous studies have reported time-reversal-symmetry (TRS) breaking below a temperature scale that coincides with the onset of magnetoresistance. The simultaneous enhancement of both $T_{\rm c}$ and magnetoresistance under stress therefore suggests a positive correlation between superconductivity and normal-state electronic and magnetic properties in LaRu$_{3}$Si$_{2}$. Detailed calculations demonstrate that stress-induced changes in $T_{\rm c}$ arise from the joint evolution of the total density of states and the flat band, whereas the large magnetoresistance enhancement is dominated by the stress-driven downward shift of the Ru $dz^{2}$ kagome flat band.

\end{abstract}

\maketitle

Kagome superconductors have emerged as a fertile ground for investigating correlated electron behavior, owing to their geometrically frustrated lattice\cite{syozi1951statistics} and the delicate balance between competing quantum states\cite{wilson2024v3sb5,ortiz2019new,yin2022topological,guguchia2023unconventional,neupert2022charge, jiang2021unconventional,Guo2022,Christensen2022,mielke2022time,1g9n-wm38}. This competition – particularly between superconductivity and charge order – renders these systems highly tunable through external parameters such as chemical substitution or the application of mechanical pressure\cite{guguchia2023tunable,lin2024uniaxial,graham2025pressure,mielke2024microscopic,plokhikh2024discovery,li2025superconducting}.\par

Besides the deeply studied family of \textit{A}V$_{3}$Sb$_{5}$ (\textit{A} = K, Rb, Cs)\cite{wilson2024v3sb5,ortiz2019new,mielke2022time,khasanov2022time,guguchia2023tunable} showing an exceptionally rich spectrum of electronic properties including unconventional superconductivity competing with charge order for the same electronic states, LaRu$_{3}$Si$_{2}$\cite{mielke2021nodeless,mielke2025coexisting,Barz1980,Vandenberg1980,li2016chemical,deng2025theory} represents an excellent opportunity to further extend the study of entanglement between these two phenomena. Featuring kagome superconductivity below $T_{\rm c}$ ${\approx}$ 7 K  and an unusually high charge-ordering temperature ($T_{\rm co,I}$ ${\approx}$ 400 K), followed by the emergence of a secondary charge order below $T_{\rm co,II}$ ${\approx}$ 80 K and a weak magnetic state below $T^{*}$ ${\approx}$ 35 K accompanied by large magnetoresistance, this system provides a unique platform for exploring the intricate interplay among intertwined quantum states in kagome materials. Related compounds \cite{kral2026chargeorder,https://doi.org/10.1002/adma.202513015,gaggl2025bulksuperconductivitykagomemetal} from the same structural family have recently been shown to exhibit similarly rich physics, thereby bringing the 132 family of kagome superconductors into greater prominence and making it an increasingly attractive platform for investigating correlated quantum phenomena.\par

\begin{figure*}
    \centering
    \includegraphics[width=1.0\linewidth]{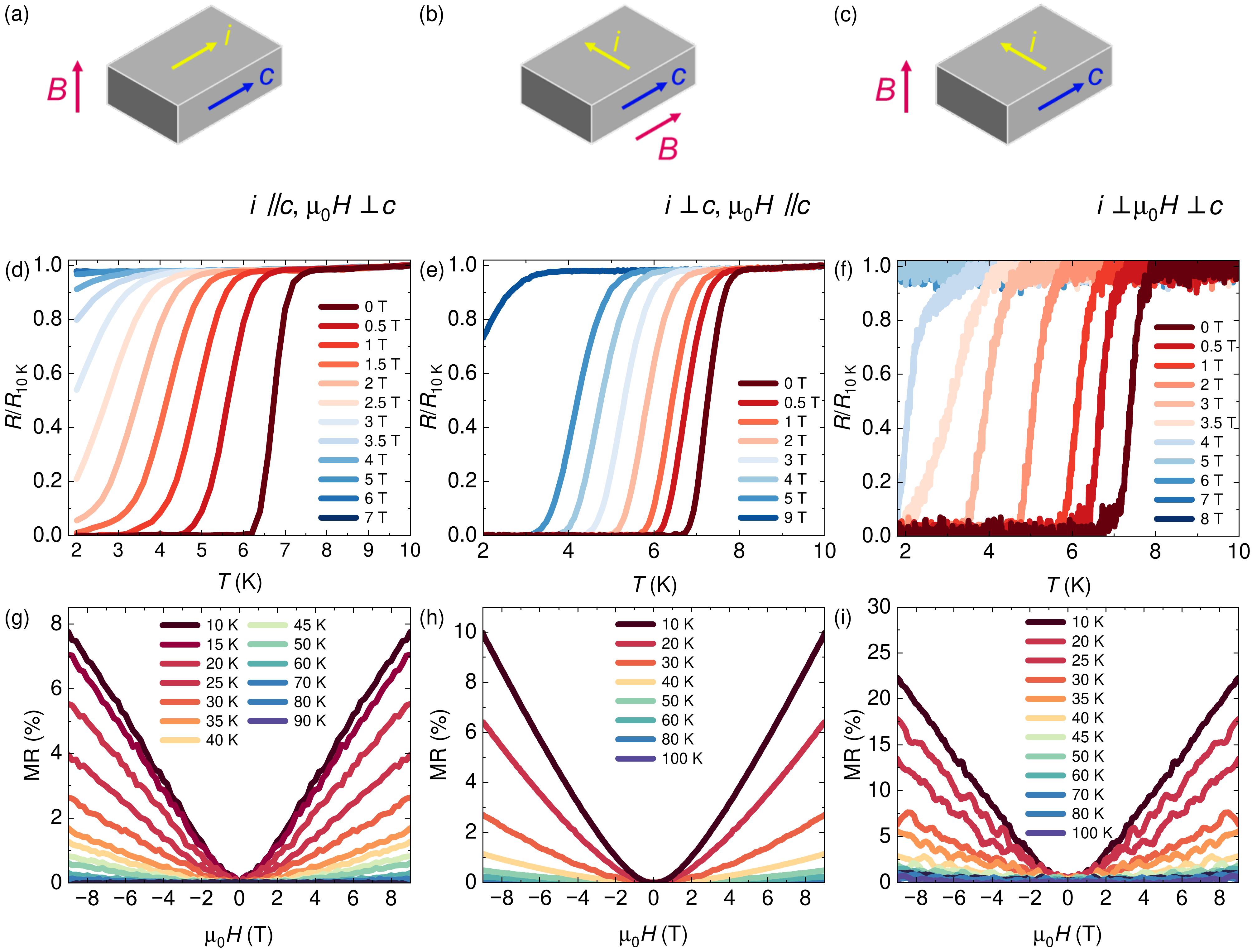}
    \caption{(a-c) Schematic representations of the experimental configurations with varying orientations of the electrical current and magnetic field relative to the crystallographic 
$c$-axis. Panels below display the corresponding measurements for each configuration. (d-f) Electrical resistivity across the superconducting transition measured under the application of various magnetic fields. (g-i) Magnetoresistance curves measured at selected temperatures for each of the three measurement configurations.}
    \label{fig:ambient}
\end{figure*}

Recent high-pressure studies on LaRu$_{3}$Si$_{2}$ \cite{ma2025correlation,li2025superconducting} reveal a dome-shaped superconducting phase diagram with an optimal transition temperature of $T_{\rm c}\approx9$ K, the highest reported amongst kagome superconductors. This evolution of $T_{\rm c}$ appears to originate from an intricate interplay between superconductivity, charge order, and the underlying normal-state electronic response. In particular, pressure uncovers a clear positive correlation between superconductivity and normal-state electronic features associated with the characteristic temperature scale $T^{*}$. Moreover, pressure studies demonstrate a strong correlation between $T_{\rm c}$ and the spatial coherence of the charge order: as long as the charge order remains long-ranged,  $T_{\rm c}$ reaches its optimal value, whereas a pressure-induced crossover from long- to short-range charge order leads to a substantial suppression of $T_{\rm c}$. These findings highlight the importance not only of the presence of charge order, but also of its spatial character and its coupling to the electronic and magnetic landscape of the normal state. In LaRu$_{3}$Si$_{2}$, the charge order is intimately linked to distortions within the Ru layer, from which time-reversal-symmetry (TRS) breaking also emerges\cite{plokhikh2024discovery,mielke2025coexisting}. This strongly suggests that the kagome lattice plays a central role in governing superconductivity, charge order, and TRS breaking, providing a natural explanation for the observed direct correlation between superconductivity and charge order. A key next step is therefore to identify external tuning parameters that can optimize the Ru-site distortions and the associated electronic responses to further enhance $T_{\rm c}$. In this context, applying uniaxial stress along multiple crystallographic directions offers a particularly promising route to selectively tune Ru-site distortions, $T^{*}$, and $T_{\rm co,II}$, with the potential to further increase the superconducting transition temperature. Uniaxial stress has been shown to be an efficient tool to tune the superconductivity and magnetism in cuprates\cite{guguchia2020using,guguchia2024designing,thomarat2024tuning,islam2025contrasting} and Sr$_{2}$RuO$_{4}$\cite{grinenko2021split,grinenko2023mu}, however, the effect of uniaxial compression is much less explored in the family of kagome materials\cite{lin2024uniaxial,sadrollahi2025straincontrolelectronicsuperlatticedomains,PhysRevB.104.144506,gerguri2025distinct}.\par

In this work, we first investigate the anisotropic electronic response of LaRu$_{3}$Si$_{2}$ using detailed magnetotransport measurements. By systematically varying the orientations of the electrical current and magnetic field with respect to the crystallographic $c$-axis, we reveal a pronounced directional dependence in both the normal and superconducting states. Building on these results, we then explore the effects of in-plane uniaxial stress, focusing on the configuration in which the intrinsic anisotropy already enhances the normal-state response. Upon applying in-plane uniaxial stress of up to 0.6 GPa, we observe a modest increase in the superconducting transition temperature $T_{\rm c}$ together with a substantial enhancement of the magnetoresistance. The concurrent enhancement of superconducting and normal-state electronic responses under stress, linked to changes in the kagome electronic structure, highlights a positive correlation between superconductivity and the underlying electronic and magnetic properties.\par

\begin{figure*}
    \centering
    \includegraphics[width=1.0\linewidth]{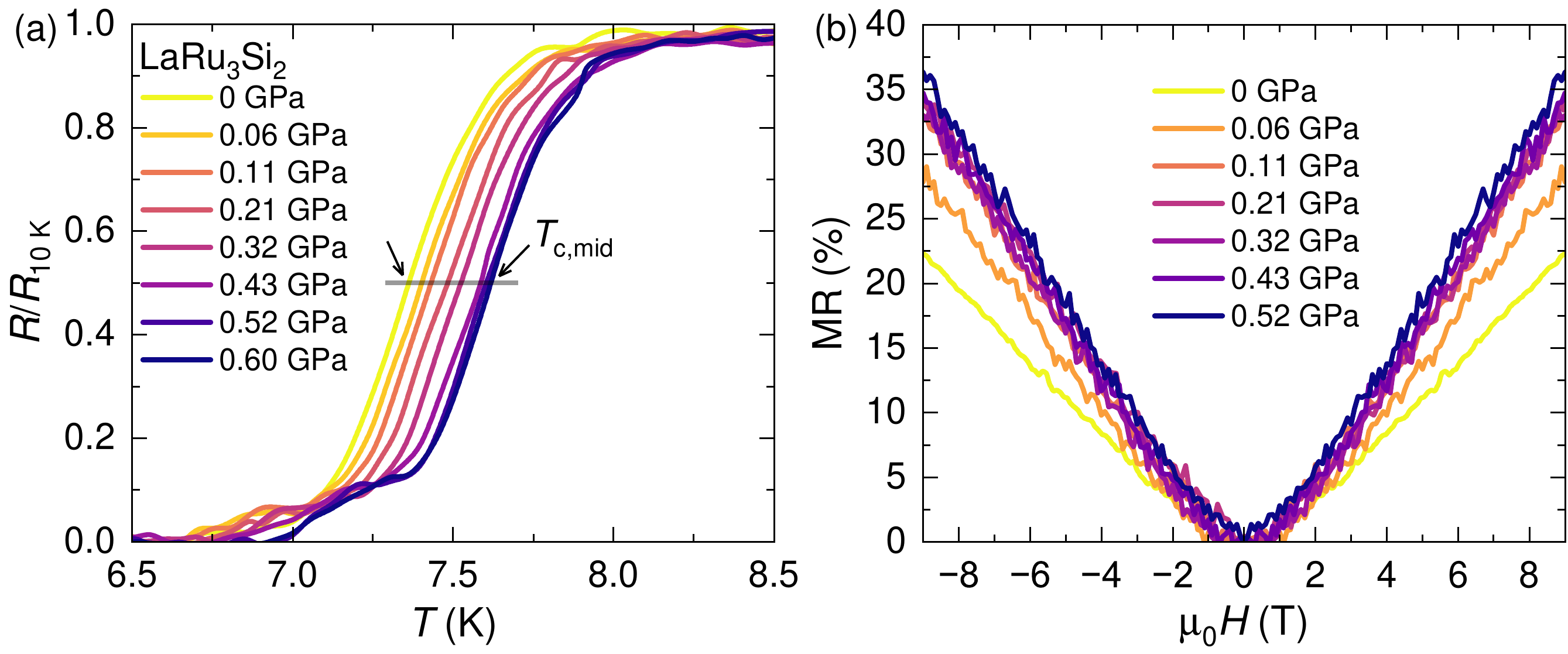}
    \caption{(a) The temperature dependence of electrical resistivity, normalized to its value at 10 K, for LaRu$_{3}$Si$_{2}$, measured under various in-plane uniaxial stress values. (b) The field dependence of magnetoresistance, measured under various in-plane uniaxial stress values. These measurements were performed with the following configuration $i \perp \mu_{0}H \perp c$.}
    \label{fig:strain}
\end{figure*}

Figure \ref{fig:ambient}  summarizes the magnetotransport measurements of LaRu$_{3}$Si$_{2}$ in both the superconducting and normal states, obtained for different orientations of the electrical current and magnetic field with respect to the crystallographic $c$-axis. The three experimental configurations investigated are illustrated in Figs. \ref{fig:ambient}(a-c). The temperature dependence of the resistivity across the superconducting transition for these configurations is shown in Figs. \ref{fig:ambient}(d-f). Well-defined superconducting transitions are observed in all cases, with a midpoint transition temperature of approximately 7 K. Notably, the upper critical field $\mu_{0}H_{\rm c2}$ exhibits a pronounced anisotropy, varying significantly among the three configurations. For the two configurations with the magnetic field in the Kagome plane, $i \parallel c$, $\mu_{0}H \perp c$ and $i \perp \mu_{0}H \perp c$, investigated in the present work, the upper critical field is nearly identical, with $\mu_{0}H_{\rm c2}$ ${\simeq}$ 4–5 T at 2 K. A slightly higher value of 5 T has been previously reported for the configuration $i \parallel \mu_{0}H \parallel c$ \cite{mielke2024charge}. In contrast, for the configuration $i \perp c$, $\mu_{0}H \parallel c$, the superconductivity is significantly more robust, with the upper critical field exceeding $\mu_{0}H_{\rm c2}$ ${\textgreater}$ 9 T at 2 K. These results demonstrate that the relative orientation of the electrical current and magnetic field plays a decisive role in determining the magnitude of $\mu_{0}H_{\rm c2}$. The upper critical field $\mu_{0}H_{\rm c2}$ is governed by Cooper-pair breaking mechanisms, which can be classified as either orbital pair breaking—arising from vortex formation due to the Lorentz force—or Pauli (Zeeman) pair breaking, originating from spin polarization. Orbital pair breaking is inherently anisotropic and depends sensitively on the orientation of the applied magnetic field relative to the crystallographic axes and the Fermi-surface geometry, whereas Pauli limiting is largely isotropic unless strong spin–orbit coupling is present. For magnetic fields applied parallel to the crystallographic $c$-axis  ($H \parallel c$), vortices penetrate perpendicular to the kagome planes. In this configuration, the relevant orbital length scale is set by the in-plane superconducting coherence lengths, which are comparatively large, resulting in weaker orbital pair breaking and consequently a larger upper critical field $\mu_{0}H_{\rm c2}$. This behavior is characteristic of layered or quasi-two-dimensional superconductors.
Taken together, these observations demonstrate that superconductivity in LaRu$_{3}$Si$_{2}$ is strongly anisotropic despite its three-dimensional crystal structure. The superconducting state is dominated by the kagome planes, and the magnitude of $\mu_{0}H_{\rm c2}$ is primarily controlled by the orientation of the magnetic field rather than the direction of the electrical current. Moreover, the data indicate that superconductivity in LaRu$_{3}$Si$_{2}$ is predominantly orbital-limited rather than Pauli-limited.

\begin{figure*}
    \centering
    \includegraphics[width=1.0\linewidth]{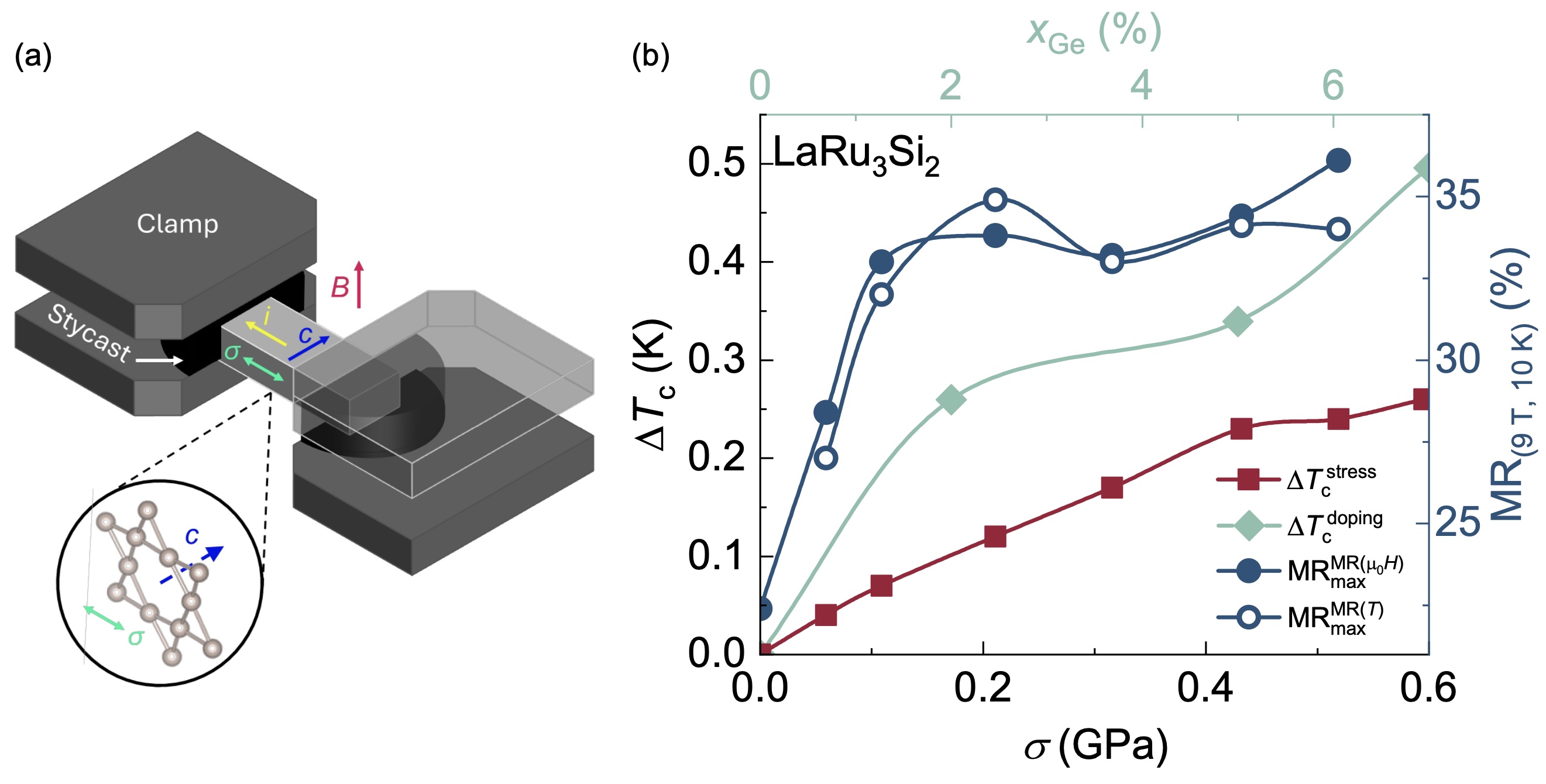}
    \caption{(a) A simplified schematic illustration of the uniaxial stress setup, with the directions of the applied stress, transport current, and magnetic field indicated. (b) Uniaxial stress dependence of the superconducting transition temperature (left axis) and the magnetoresistance (right axis) in LaRu$_{3}$Si$_{2}$. The chemical doping (LaRu$_{3}$(Si$_{1-x}$Ge$_{x}$)$_{2}$) effect on $T_{\rm c}$ is shown for comparison, as well\cite{misawa2025chemical}.}
    \label{fig:summary}
\end{figure*}

To further elucidate the anisotropic behavior, we focus on the normal-state response that is closely linked to superconductivity \cite{ma2025correlation}. In particular the magnetoresistance, which reflects the onset of secondary charge-order formation in LaRu$_{3}$Si$_{2}$ below $T_{\rm co,II}$ ${\approx}$ 80 K and the emergence of a weak magnetic state below $T^{*}$ ${\approx}$ 35 K. Magnetoresistance curves measured at selected temperatures are shown in Figs. \ref{fig:ambient}(g–i). Independent of the measurement configuration, the magnetoresistance onset is observed below 80 K, followed by a pronounced enhancement at temperatures below 35 K. Notably, the relative orientation of the magnetic field and electrical current has a strong impact on the magnitude of the magnetoresistance. In the configurations $i \parallel c$, $\mu_{0}H \perp c$, and $i \parallel \mu_{0}H \parallel c$ (reported previously), a moderate magnetoresistance of approximately 8 ${\%}$ is observed at $T$ = 10 K and $\mu_{0}H$ = 9 T. For the configuration $i \perp c$, $\mu_{0}H \parallel c$, the magnetoresistance increases to about 10 ${\%}$. In contrast, for $i \perp \mu_{0}H \perp c$, the magnetoresistance is strongly enhanced, reaching nearly 23 ${\%}$, exceeding the maximum value of 14 ${\%}$ reported under hydrostatic pressure \cite{ma2025correlation}. These results further substantiate the pronounced anisotropy of both the superconducting and normal-state electronic properties of LaRu$_{3}$Si$_{2}$ and highlight a close connection between superconductivity and the normal-state electronic response, underscoring the electronic origin of these intertwined phases.\par

Focusing on the configuration with the strongest magnetotransport response (see \ref{fig:summary}a), we further aim to explore the subtle interplay between superconductivity and the normal-state electronic properties through in-plane uniaxial stress. Fig. \ref{fig:strain}(a) depicts the stress effect on the superconducting transition, clearly indicating the gradual increase in all the onset, midpoint, and zero-resistance characteristic temperatures setting the in-plane uniaxial stress as an effective tool for optimizing the superconducting properties. The stress-driven increase in $T_{\rm c}$ is accompanied by the significant enhancement of the maximum value of magnetoresistance reached, from ambient $\approx$23\% to the saturation at $\approx$35\% above $\sigma \approx$ 0.1 GPa. Fig. \ref{fig:summary}b summarizes the evolution of the superconducting transition temperature and the maximum magnetoresistance (measured at $T$ = 10 K and $\mu_{0}H$ = 9 T) as a function of uniaxial stress. The data show a gradual increase in $T_{\rm c}$ with increasing stress, reflecting a clear enhancement of the superconducting properties. The magnetoresistance follows a similar overall trend, however, initially rising rather sharply before eventually saturating at the maximum value of $\approx$35\% at higher stress applied.\par

The simultaneous enhancement of both $T_{\rm c}$ and magnetoresistance under in-plane uniaxial stress constitutes a key finding of this work. The positive stress dependence of $T_{\rm c}$ is encouraging, as it motivates extending uniaxial stress to higher values with the potential for further enhancement of the superconducting transition temperature. More strikingly, the magnetoresistance increases by approximately 60${\%}$ under a relatively small in-plane uniaxial stress of only 0.1 GPa. As discussed above, the onset of magnetoresistance coincides with the emergence of secondary charge order, while its strongest enhancement occurs below 35 K, the temperature at which muon-spin rotation measurements reveal the onset of weak magnetism. The pronounced stress-induced enhancement of MR therefore points to a modification of the charge-ordered state and the associated electronic structure, an enhancement of the magnetic response, or a combined effect of both. The strong sensitivity of the normal-state electronic and magnetic properties to in-plane uniaxial stress makes LaRu$_{3}$Si$_{2}$ particularly intriguing from both fundamental and potential application perspectives. Furthermore, the observed positive correlation between $T_{\rm c}$ and magnetoresistance highlights a strong interdependence between superconductivity and normal-state electronic or magnetic properties, suggesting a common underlying mechanism. Taken together, these results strongly support an electronically driven superconducting state in LaRu$_{3}$Si$_{2}$.\par

\begin{figure*}[htp!]
	\centerline{\includegraphics[width = 1\linewidth]{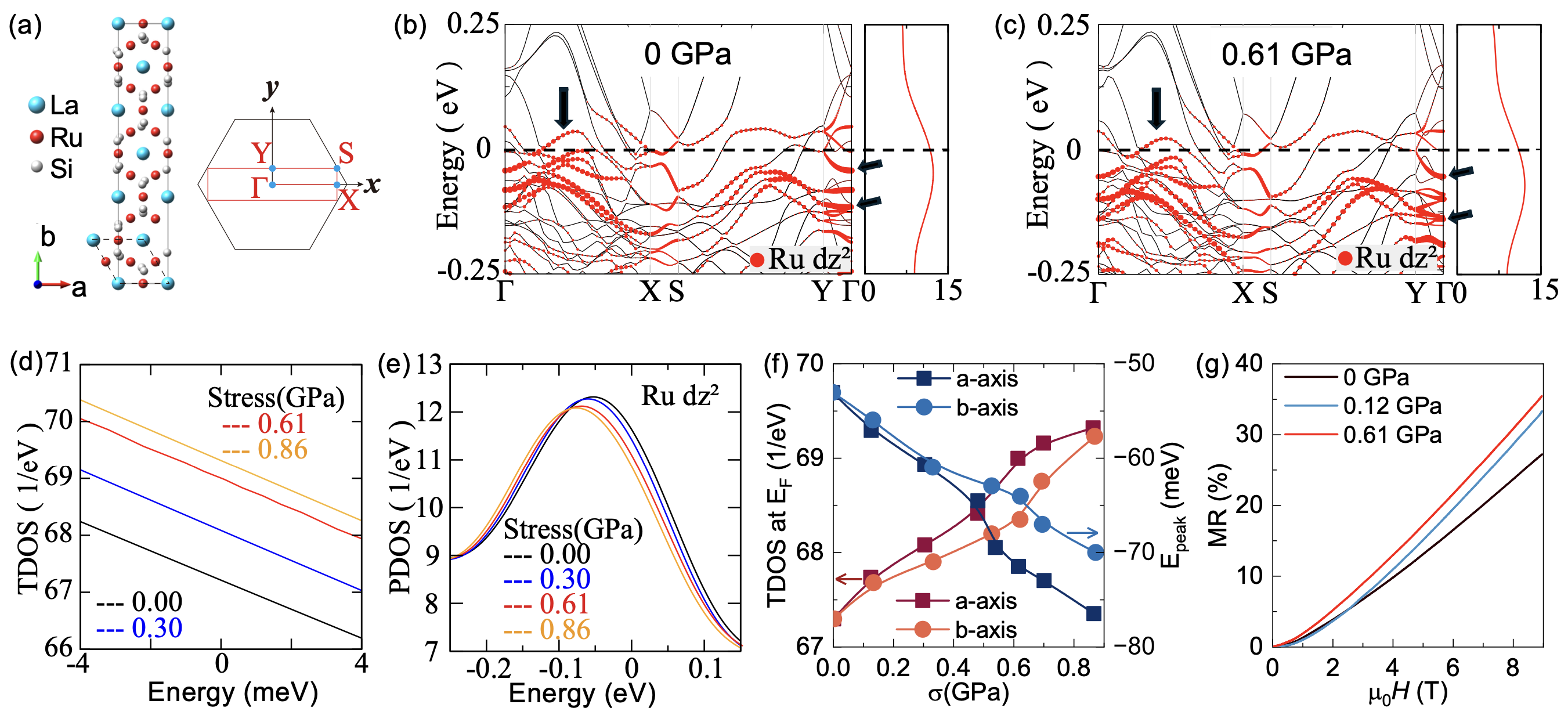}}
	\caption{\textbf{(a)} Crystal structure and Brillouin zone ($k_z=0$ plane) of the CO-II phase, with the folded Brillouin zone induced by structural distortion indicated. \textbf{(b, c)} Band structures and projected density of states (PDOS) at 0~GPa \textbf{(b)} and under 0.61~GPa uniaxial stress along the $a$-axis \textbf{(c)}. The size of the red circles scales with the spectral weight of the Ru $d_{z^2}$ orbitals. Dashed lines and arrows are intended to guide the reader in identifying stress-induced changes. The dashed lines indicate the Fermi level, while the arrows highlight regions showing a continuous shift of the bands away from the Fermi level. \textbf{(d)} Evolution of the total density of states (TDOS) near the Fermi level under uniaxial stress along the $a$-axis.  \textbf{(e)} Evolution of the Ru $d_{z^2}$ PDOS as a function of uniaxial stress.  \textbf{(f)} (left axis) Calculated TDOS at the Fermi level as a function of uniaxial stress applied along the $a$ and $b$ axes. (right axis) Energy position of the Ru $d_{z^2}$-dominated flat-band peak center as a function of uniaxial stress along the $a$ and $b$ axes. \textbf{(g)} Calculated magnetoresistance as a function of magnetic field $H$ for different $a$-axis uniaxial stress values.}
	\label{fig:structure_bands}
\end{figure*}

To elucidate the microscopic origin of the uniaxial-stress effects, we performed first-principles calculations as a function of applied stress. The intrinsic structural distortion of the Ru kagome sublattice leads to a reconstruction of the Brillouin zone (BZ), as illustrated in Fig. \ref{fig:structure_bands}a. The calculated band structures and projected density of states (PDOS) of the Ru $d_{z^2}$ orbitals at 0 GPa and under 0.61 GPa uniaxial stress applied along the \textit{a}-axis are shown in Fig. \ref{fig:structure_bands}b and c, respectively. As shown in Fig. \ref{fig:structure_bands}b, some quasi-flat bands close to the Fermi level are observed in the whole reconstructed BZ. These bands are dominated by Ru $d_{z^{2}}$ orbitals and obtained by the folding effect of kagome flat band for the SG 193 structure  in previous reports~\cite{PhysRevMaterials.5.034803}.
From these calculations, we extracted several key quantities: 
(i) the evolution of the total density of states (TDOS) near the Fermi level under uniaxial stress (Fig. \ref{fig:structure_bands}d); (ii) the evolution of the Ru $d_{z^{2}}$ flat bands and PDOS under uniaxial stress (Fig. \ref{fig:structure_bands}e); (iii) the calculated magnetoresistance as a function of magnetic field and $a$-axis uniaxial stress (Fig. \ref{fig:structure_bands}g). With increasing \textit{a}-axis uniaxial stress, the electronic structure evolves with two characteristic trends (Fig. \ref{fig:structure_bands}f): a monotonic increase in the total TDOS (Fig. \ref{fig:structure_bands}f left axis) and a concomitant downward shift of the Ru $d_{z^2}$ quasi-flat band in energy (Fig. \ref{fig:structure_bands}f right axis). A qualitatively similar behavior is found for stress applied along the \textit{b}-axis.

In the following, we discuss how the theoretical results
align with the experimental observations, namely
a modest stress-induced enhancement of Tc and a pronounced
increase in magnetoresistance. The calculated
enhancement of the TDOS at the Fermi level (Fig. \ref{fig:structure_bands}d
and f) can contribute to the observed increase in T$_c$. In
contrast, the stress-induced shift of the Ru $d_{z^{2}}$ PDOS
peak (Fig. \ref{fig:structure_bands}e and f) indicates that the kagome-derived
flat band moves away from E$_F$ . This shift is expected to
weaken electronic correlations and reduce the flat-band
contribution to superconducting pairing, which would
otherwise favor a higher T$_c$. As a consequence, two competing
effects are at play: the increase in TDOS tends to
enhance T$_c$, while the reduction of flat-band correlations
counteracts this tendency. The resulting modest net increase
in T$_c$ therefore likely reflects the displacement of
the flat band away from the Fermi level, which limits the
overall enhancement of superconductivity.
Importantly, the same stress-induced shift of the flat band has a much stronger impact on the normal-state transport. As the flat band moves away from E$_F$, the heavy carriers contribution to the Fermi surface is reduced, leading to a decrease in the average quasiparticle effective mass m$^\ast$. Within the Chambers transport formalism~\cite{ashcroft1976solid}, a reduction in the effective mass enhances the cyclotron frequency, resulting in more rapid deflection of carriers and thereby tending to enhance the magnetoresistance. To validate this physical picture, we calculated the MR using a tight-binding model based on SG 193 structure, since it shares similar distorted Ru-kagome layers with the CO-II phase and exhibits a consistent pressure-driven evolution of the TDOS and PDOS. Our calculations (Fig. \ref{fig:structure_bands}g) successfully reproduce the experimentally observed large enhancement of the magnetoresistance. Furthermore, the calculations reveal a pronounced
initial increase of the magnetoresistance (MR) between 0 and 0.12 GPa, followed by a much weaker enhancement at higher applied stress. 
This behavior is in excellent agreement with the experimental observations, further substantiating the consistency between experiment and theory.
These results demonstrate that, while the stress-induced change in Tc arises from a combined evolution of the TDOS and the flat band, the dramatic increase in magnetoresistance can be explained predominantly—if not entirely—by the stress-driven evolution of the kagome flat band.

An increase of $T_{\rm c}$ in LaRu$_{3}$Si$_{2}$ upon substituting Si with Ge has been reported recently \cite{misawa2025chemical}. This behavior contrasts sharply with chemical substitution at the Ru site, which generally leads to a pronounced suppression of superconductivity \cite{li2012distinct,li2016chemical}, underscoring a fundamental difference between the substitution mechanisms. Replacing Si with the larger Ge atom induces a strongly anisotropic lattice expansion, dominated by an elongation along the crystallographic $c$-axis. This response is analogous to the effect of in-plane uniaxial stress, which likewise causes $c$-axis expansion. 
However, the mechanical tuning employed in this work provides a cleaner approach that preserves the intrinsic chemical structure, enabling, for the first time, a direct assessment of the role of $c$-axis effects on the electronic structure, superconductivity, and magnetoresistance.

In conclusion, we combine magnetotransport measurements and density-functional-theory (DFT) calculations to investigate the superconducting and normal-state properties of LaRu$_{3}$Si$_{2}$ under in-plane uniaxial stress. We first identify a pronounced intrinsic anisotropy in both superconducting and normal-state responses, with the upper critical field and the magnitude of the magnetoresistance strongly dependent on the relative orientations of the electrical current and magnetic field with respect to the crystallographic $c$-axis. The magnetoresistance reaches its maximum when both the electrical current and the magnetic field are oriented perpendicular to the $c$-axis. The strong dependence of upper critical field and MR on magnetic-field orientation, despite identical current directions, demonstrates that superconductivity in LaRu$_{3}$Si$_{2}$ is governed by anisotropic orbital pair breaking associated with the kagome planes.  Focusing on the configuration, which maximizes the normal-state MR response, we apply in-plane uniaxial stress up to 0.6 GPa and observe a monotonic increase of the superconducting transition temperature  $T_{\rm c}$. Although the enhancement of  $T_{\rm c}$ is modest, the positive pressure dependence is notable. In contrast, the magnetoresistance exhibits a giant enhancement of up to 60$\%$ under in-plane compression. Our calculations indicate that the stress-induced change in $T_{\rm c}$ reflects the combined evolution of the TDOS and the flat band, while the large magnetoresistance enhancement is dominated by the stress-driven evolution of the kagome flat band. Given that the onset of weak magnetism coincides with the emergence of magnetoresistance, these results further suggest a stress-induced modification of the magnetic response. Taken together, our findings reveal a positive coupling between superconductivity and normal-state electronic properties, highlight the central role of the kagome flat band in both the normal and superconducting states, and point to an electronically driven superconducting state in LaRu$_{3}$Si$_{2}$.

\section*{Methods}

\textbf{Uniaxial Stress Cell:} For the uniaxial stress experiment, we employed the FC100 stress cell (Razorbill Instruments) compatible with the Physical Property Measurement System (PPMS, Quantum Design). A single-crystalline LaRu$_3$Si$_2$ sample was mounted between two piezoelectrically actuated plates using Stycast epoxy. Magnetotransport measurements were performed using the standard four-probe technique. The experimental geometry was configured according to Fig. \ref{fig:ambient}(c), i.e., $i \perp \mu_{0}H \perp c$, with the stress direction parallel to the electrical current.\par
Razorbill FC100 is a force-controlled stress cell, allowing direct measurement of the force applied on the sample. Knowing the sample width and thickness (0.9 and 0.4 mm, in this case), the corresponding stress can be determined. Within the experiment performed, the maximum force of 214 N was applied, corresponding to the stress of 0.60 GPa.\\

\textbf{Computational Method:} 
The first-principles calculations were performed using the Vienna Ab initio Simulation Package (VASP) based on the density functional theory (DFT)~\cite{PhysRevB.50.17953,PhysRevB.54.11169,kresse1999ultrasoft}. The exchange-correlation functional was treated using the generalized gradient approximation (GGA) parameterized by Perdew, Burke, and Ernzerhof (PBE) ~\cite{PhysRevLett.77.3865}. The cutoff energy for wave function expansion is 380 eV, and a $9 \times 3 \times 7$ Monkhorst-Pack $k$-point mesh was employed for the Brillouin zone sampling. 
To simulate the effects of in-plane uniaxial stress, the CO-II phase of LaRu$_3$Si$_2$ (SG 55) was optimized with the lattice parameter constrained along the applied stress direction, while allowing the remaining lattice parameters and all internal atomic positions to fully relax until the residual forces on each atom were less than 0.01 eV/\AA. 
For the transport calculations, a tight-binding Hamiltonian based on the room-temperature structure (SG 193) was constructed using maximally localized Wannier functions (MLWFs) via the Wannier90 package~\cite{PhysRevB.56.12847,PhysRevB.65.035109,RevModPhys.84.1419,MOSTOFI20142309}. Subsequently, magnetoresistance calculations were performed using the WannierTools package~\cite{WU2018405,PhysRevB.99.035142}, taking into account the flat-band-induced quasiparticle mass renormalization.

\subsection*{Data availability}
All related data are available from the authors on request.

\subsection*{Acknowledgments}
This work is based on experiments performed at the Swiss Muon Source (S${\mu}$S) Paul Scherrer Insitute, Villigen, Switzerland. Z.G. acknowledges support from the Swiss National Science Foundation (SNSF) through SNSF Starting Grant (No. TMSGI2${\_}$211750) and through Mouniverse NCCR 51AU40${\_}$1229254. Z.G. and H.L. acknowledge useful discussion and technical support from Dr. Marisa Medarde. The National Key Research and Development Program of China (2024YFA1611200), and the National Natural Science Foundation of China (Grant No. 12274154). The computations were completed in the HPC Platform of Huazhong University of Science and Technology.\\

\section{Author contributions}~
Z.G. conceived, designed and supervised the project. Sample growth: H.K. and S.N.. Magnetotransport experiments as a function of uniaxial stress: Z.G., P.K., V.S., O.G., M.S., J.N.G. with contribution from M.B.. DFT calculations: Y.G. and G.X. in consultation with Z.G. Figure development and writing of the paper: Z.G. and P.K.. All authors discussed the results, interpretation, and conclusion.\\ 

\section*{Data availability}
The data that support the findings of this study are available from the corresponding authors upon request.\\

\section*{Conflict of Interest}
The authors declare no financial/commercial conflict of interest.\\

\bibliography{References}{}

\end{document}